\documentclass[twocolumn,showpacs,preprintnumbers,amsmath,amssymb,pra, superscriptaddress,epsfig]{revtex4}
\usepackage{graphicx}% Include figure files
\usepackage{dcolumn}% Align table columns on decimal point
\usepackage{bm}% bold math
\usepackage{color}%color text for editing}

\def\fig_width{3. in} % width of single column figure in PR

\begin{document}

\title{Measurement of hyperfine structure and isotope shifts in the Dy 421-nm transition}
%%%----------------------------------------------------------------------
\author{N. Leefer}
\email{naleefer@berkeley.edu} \affiliation{Department of
Physics, University of California at Berkeley, Berkeley,
California 94720-7300, USA}
\author{A. Cing{\"{o}}z}
\email{acingoz@berkeley.edu} \affiliation{Department of
Physics, University of California at Berkeley, Berkeley,
California 94720-7300, USA}
\author{D. Budker}
\email{budker@berkeley.edu} \affiliation{Department of
Physics, University of California at Berkeley, Berkeley,
California 94720-7300, USA} \affiliation{Nuclear Science Division,
Lawrence Berkeley National Laboratory, Berkeley, California 94720,
USA}

\date{\today}

%%%----------------------------------------------------------------------

%\doublespacing

\begin{abstract}
A measurement of the hyperfine coefficients and isotope shifts for the Dy I 421.291 nm transition [$4f^{10}6s^2\, (J=8)\rightarrow4f^{10}6s6p\, (J=9)$] using atomic beam laser-induced fluorescence spectroscopy is presented. A King Plot analysis is performed to determine a specific mass shift of $\delta \nu_{sms}^{164-162}=11(7)$~Hz for the 421-nm transition, confirming the pure $4f^{10}6s6p$ configuration of the excited state.  This transition is currently being explored for laser cooling of an atomic beam of dysprosium used in a search for a temporal variation of the fine-structure constant.
\end{abstract}

%\doublespacing

\maketitle

\section{Introduction}

Recent progress in cooling and trapping atoms with large magnetic moments presents numerous opportunities for new studies in areas such as degenerate dipolar Fermi gases (DDFG)~\cite{Lewenstein2000,Lev2009}, deterministic single atom sources~\cite{McClelland2003}, and quantum information~\cite{Lewenstein2008}.  Among the possible candidates for these studies, dysprosium (Dy) presents itself as an ideal case with one of the largest ground state magnetic moments in the periodic table ($\sim$~10 $\mu_B$).  In addition to the large ground state magnetic moment there is a pair of nearly degenerate excited state energy leves~\cite{BudkerThesis}, presenting the opportunity for fundamental-physics tests~\cite{Nguyen1997,Nguyen2004}.  Our group is currently exploring the possibility of laser cooling atomic dysprosium (Dy, atomic number $Z$=66) using a strong cycling transition at 421-nm~\cite{Leefer2008}, for which detailed knowledge of the hyperfine and isotopic structure of the transition is required.  While our current motivation is the transverse cooling of a thermal atomic beam used in a search for variation of the fine structure constant~\cite{Cingoz2007}, future interest is in the development of a magneto-optical trap (MOT) version of the experiment.  As of today, little information about this cooling transition is available in the literature despite it being the strongest line in the visible spectrum of dysprosium.  In this paper, we present the results of our measurement of the hyperfine coefficients and isotope shifts for the transition.

\section{Experimental Setup}

\subsection{Apparatus}

Spectroscopy was performed on a thermal atomic-beam using a beam source previously used in a search for a temporal variation of the fine-structure constant~\cite{Cingoz2007}.  A detailed description of the atomic-beam source is given in Ref.~\cite{Nguyen1997}. The beam is produced by an effusive oven with a multislit nozzle-array operating at $\simeq$~1500~K. The oven consists of a molybdenum tube containing dysprosium metal, and is surrounded by resistive heaters made from tantalum wire enclosed inside alumina ceramic tubes. The atomic beam has a mean velocity of $\simeq 5\times 10^4$ cm/s with a full-angle divergence of $\simeq 0.2$ rad (1/$e^2$ level) in both transverse directions.

To generate the 421-nm light, approximately 15~W of Ar-ion laser light (Coherent Innova 400) was used to pump a Ti:Sapphire ring laser (Coherent 899), producing up to 650 mW of 842-nm light.  After diagnostics and passing through an optical fiber approximately 250 mW of 842-nm light made a single pass through a 1$\times$2$\times$10 mm periodically poled potassium titanyl phosphate (PPKTP) crystal (Raicol Crystals Ltd.), producing $\sim300$~$\mu$W of 421-nm light.  The 421-nm light entered the atomic-beam apparatus and intersected the atoms at a ninety-degree angle.

The atomic fluorescence emitted perpendicular to both the atomic and laser beams was detected with a photomultiplier tube (PMT) through an optical viewport with a 420-nm interference filter ($\sim48$\% peak transmission, 10-nm bandwidth) at the input window of the PMT.  The PMT signal was digitized with a 12-bit oscilloscope (Yokagawa DL9040).  The fluorescence signal was not normalized to the 421-nm power as each scan over the spectrum was short ($<$~1 s) and power fluctuations were verified to be negligible by monitoring transmission amplitude through a Fabry-Perot (FP) cavity.  There was a change in 421-nm power caused by the changing efficiency of second-harmonic generation (SHG) as the laser frequency was scanned. However, this change in power was insignificant over the width of the transition ($\sim$3 GHz in the Ti:Saph scan).

\subsection{Frequency Calibration}

To accurately determine frequency intervals we used the method of modulation spectroscopy~\cite{Wijngaarden}, where an electro-optic modulator (EOM) was used to create spectral sidebands on the laser light exciting the atoms.  In our setup the 421-nm light made a single pass through an EOM operating at $\nu_m = $~8.870 GHz, producing sidebands on the laser at $\nu_{\mathrm{421}}~\pm\,n\,\nu_m$ ($n = 1,2,3,\cdots$).  These sidebands generate multiple copies of the fluorescence spectrum and allow for accurate calibration of frequency intervals in the laser scan.  The modulation frequency was generated by doubling the output of an rf-synthesizer (IFR 2042) locked to a cesium frequency standard (HP 5061).  Nonlinearities in the laser scan were corrected for by detecting transmission of the 842-nm laser light through a FP cavity while the laser was scanning.  It was verified that thermal drifts of the FP cavity length occurred on a longer time scale than the duration of the laser scan.

\section{Results}

\begin{figure}[t]
\includegraphics[width=3.4in]{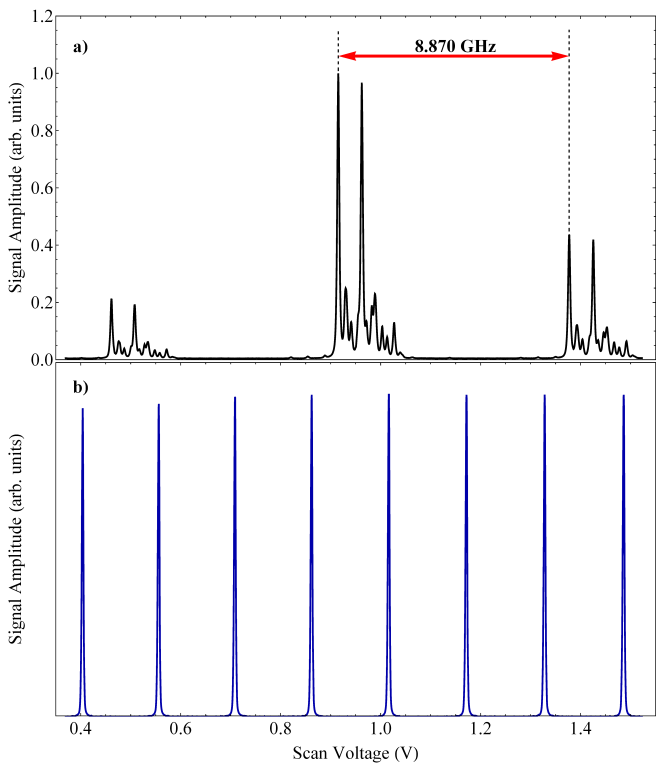}
\caption{\label{fig:Spectrum} [Color online] Example scan over the 421-nm transition. a) Fluorescence spectrum acquired with a PMT. The discrepancy in sideband amplitudes is caused by a change in second-harmonic generation (SHG) efficiency as the laser is scanned significantly from the phase matched frequency. b) The simultaneously acquired transmission signal of the 842-nm laser through a FP cavity.}
\end{figure}

An example of the data is shown in Fig.~\ref{fig:Spectrum}. Each file consisted of either a single scan or an average of 5-10 scans over the lineshape, with each scan taking $\sim$~1 second.  No significant difference was observed between averaging and not averaging. The spectrum consists of peaks corresponding to the individual hyperfine and isotopic components with a width of $\simeq$~70~MHz dominated by the residual Doppler width (the natural width of the transition is $\simeq$~30~MHz~\cite{Curry}).  Dysprosium has two odd-neutron-number isotopes ($^{161}$Dy and $^{163}$Dy with natural abundances of 19\% and 25\%, respectively) both with nonzero nuclear spin $I=5/2$.  The coupling of the nuclear spin to the total electronic angular momentum leads to a hyperfine splitting of the transition where the energy shift can be calculated according to
%\begin{table*}[t]
\begin{equation*}
\delta E_{hfs}= A\,\mathbf{I}\cdot\mathbf{J}
+B\frac{\frac{3}{2}\mathbf{I}\cdot\mathbf{J}(2\mathbf{I}\cdot \mathbf{J}+1)-I(I+1)J(J+1)}{2I(2I-1)J(2J-1)},%\\
% &+C\frac{5}{4}\frac{\{8(\mathbf{I}\cdot \mathbf{J})^3+16(\mathbf{I}\cdot \mathbf{J})^2+\frac{8}{5}(\mathbf{I}\cdot %\mathbf{J})[-3I(I+1)J(J+1)+I(I+1)+J(J+1)+3]-4I(I+1)J(J+1)\}}{I(I-1)(2I-1)J(J-1)(2J-1)}\nonumber
\end{equation*}
%\end{table*}
where $\mathbf{I}\cdot \mathbf{J}=\frac{1}{2}[F(F+1)-I(I+1)-J(J+1)]$ and $J$ and $F$ are the total electronic angular momentum and total atomic angular momentum~\cite{Childs}.  The coefficients A and B are the magnetic-dipole and electric-quadrupole hyperfine-interaction constants, respectively. The ground-state hyperfine-coefficients are available in the literature~\cite{Childs}.

\begin{figure*}[t]
\includegraphics[width=7in]{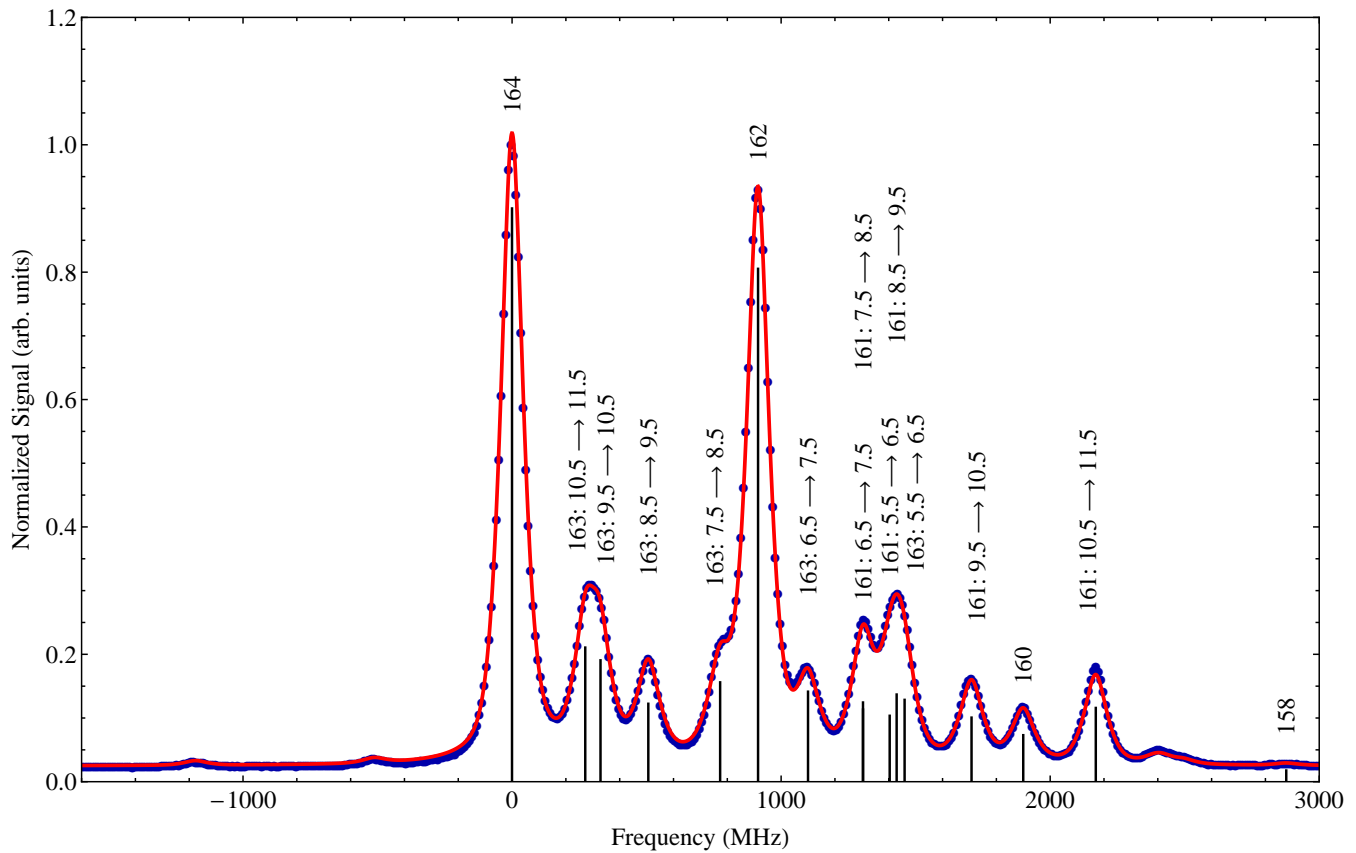}
\caption{\label{fig:ExampleFit} [Color online] Result of a nonlinear least-squares fit to a recorded spectrum.  The data are shown as circles and the fit is shown as a solid line.  The labels denote the Dy isotopes and, in the case of odd isotopes, the strongest ($F\rightarrow F+1$) hyperfine transitions are labeled.}
\end{figure*}

A nonlinear least-squares fit was performed on each spectrum, assuming a model of thirty four pseudo-Voigt profiles~\cite{Voigt} of the form
\begin{align}
\xi(\beta,\delta_i,\delta_H,\sigma,\gamma,\omega)=&\frac{\beta}{\sigma\sqrt{2\pi}}e^{-\frac{(\omega-\delta_i-\delta_H)^2}{2 \sigma^2}}\\
+&\frac{(1-\beta)\gamma}{\pi [(\omega-\delta_i-\delta_H)^2+\gamma^2]},\nonumber
\label{eqn:Voigt}
\end{align}
where $\delta_i$ and $\delta_H$ are the isotope and hyperfine shifts and $\sigma$ and $\gamma$ are the Gaussian and Lorentzian widths.  The dense structure of the transition presented difficulties in extracting isotope shifts and hyperfine coefficients without good starting values.  Initial guess values for isotope shifts were obtained by assuming a linear relationship with a change in neutron number and extracting the shifts from the 164-162 shift (the two most prominent peaks in Fig.~\ref{fig:Spectrum}).  No information was available for an initial guess of the hyperfine coefficients, except that the magnetic-dipole and electric-quadrupole coefficients were constrained so that $A^{163}/A^{161}=-1.4$ and $B^{163}/B^{161}=1.06$~\cite{Childs}, according to the ratios of the corresponding nuclear moments.  In the final fit these parameters were not constrained.  Relative peak heights were determined from isotopic abundances and from calculated relative transition amplitudes~\cite{Sobelman} for the hyperfine components of the transition.  An example fit is shown in Fig.~\ref{fig:ExampleFit}.

For each file a fit was performed on the carrier and sideband spectrums.  The relatively high laser power used ($\sim$~300~$\mu$W) was necessary to resolve isotope shifts for $^{158}$Dy and $^{156}$Dy for which isotopic abundances are small. It was found, however, that this power was enough to cause optical pumping of the hyperfine levels, making the predicted model inadequate for describing the lineshape and leading to a large systematic uncertainty in determining hyperfine coefficients and $I=5/2$ isotope shifts.  For the final results the fits performed on the smallest sideband (corresponding to the lowest power and least optical pumping) of each spectrum were used to extract hyperfine coefficients and isotope shifts for the $I=5/2$ isotopes while fits to the carrier spectrum were used to extract isotope shifts for $I=0$ isotopes.  In total 60 fits to independent lineshapes contributed to the final results presented below.  The isotope shifts and hyperfine coefficients were determined from each fit with uncertainties ranging from $0.1$ MHz to $10$ MHz depending on the relative strengths of the peaks.  The primary source of systematic error comes from uncertainties in the correction of nonlinearities in the laser scan.  These uncertainties cannot be reduced because of the large free-spectral range (FSR) of the FP cavity (1.5 GHz) relative to the width of the spectrum ($\sim$6 GHz, but this only requires a 3-GHz range for the 842-nm laser scan, see Fig.~\ref{fig:Spectrum}).

\begin{figure}[h]
\includegraphics[width=3.4in]{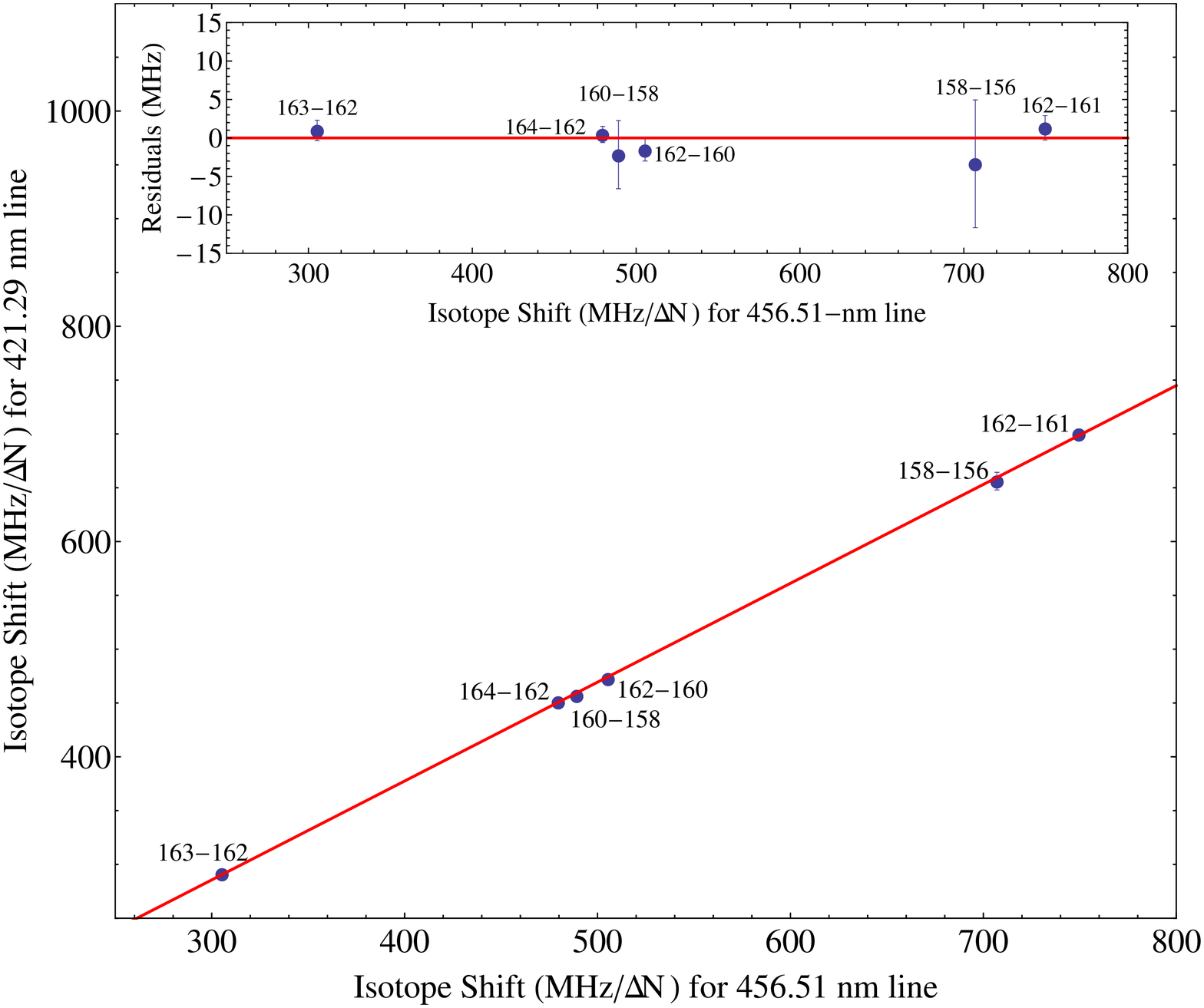}
\caption{\label{fig:KingPlot} [Color Online] Isotope shifts in the 421-nm transition vs isotope shifts in the 457-nm transition~\cite{Zaal1} normalized to the difference in nucleon number.  The solid line is a least-squares linear fit weighted by uncertainties in the x- and y-directions. The inset shows residuals between the measured isotope shifts and the fit.}
\end{figure}

\begin{table}[b]
\caption{Measured values for the hyperfine coefficients $A$ and $B$ for the $4f^{10}6s6p\, ^5K_9$ excited state ($23 736.60$ cm$^{-1}$) in dysprosium. The uncertainties quoted are one standard deviation and show an improvement of two orders of magnitude over a previous measurement~\cite{HancoxThesis}.}
\center
\begin{tabular*}{0.47\textwidth}{@{\extracolsep{\fill}} c||c|c|c|c}
\hline \hline
 & \multicolumn{2}{c}{ } \vline& \multicolumn{2}{c}{ }\\[-2ex]
 & \multicolumn{2}{c}{$A$ (MHz)} \vline& \multicolumn{2}{c}{$B$ (MHz)}  \\
\hline
 & & & & \\[-2ex]
 Isotope& This Work& Prev. Work& This Work&Prev. Work\\
\hline
 & & & & \\[-2ex]
Dy163&121.62(2)& 122(2)& 1844.9(4) & 1900(200)   \\
Dy161& -86.90(2)& -87(3)&  1747.4(5) & 1700(200) \\
\hline \hline
\end{tabular*}
\label{table:hyperfine}
\end{table}
%{0.45\textwidth}{@{\extracolsep{\fill}}

\subsection{Isotope Shifts}

\begin{table*}[t]
\caption{Isotope shifts for the 421-nm ($4f^{10}6s^2\,^5I_8\rightarrow4f^{10}6s6p\,^5K_9$) and 457-nm ($4f^{10}6s^2\,^5I_8\rightarrow4f^{10}6s6p\,^7I_8$) transitions in dysprosium.  The convention is such that $\delta \nu_{164-162}$ implies the absolute frequency of the $^{162}$Dy transition subtracted from that for the $^{164}$Dy transition.  Uncertainties are quoted for one standard deviation and all values are in MHz.}
\center
\begin{tabular*}{0.9\textwidth}{@{\extracolsep{\fill}} |l|c c c c c c| }
\hline%\\[-2ex]
\hline
&$\delta \nu_{164-163}$ &$\delta \nu_{164-162}$ &$\delta \nu_{164-161}$ &$\delta \nu_{164-160}$ &$\delta \nu_{164-158}$&$\delta \nu_{164-156}$\\
\hline
421.291 nm (this work) &-616.3(5)&-913.2(8)&-1635(1)&-1895(2)&-2868(9)&-4300(15)\\
421.291 nm (prev. work)~\cite{HancoxThesis} &-610(20)&-890(20)&-1620(20)&-1880(20)& & \\
\hline
& & & & & & \\[-2ex]
456.509 nm~\cite{Zaal1}&-660(3)&-971(2)&-1744(3)&-2020(3)&-3061(4)&-4604(5)\\
\hline
\hline
\end{tabular*}
\label{table:isotope}
\end{table*}

The measured values for isotope shifts are displayed in Table~\ref{table:isotope}, along with the isotope shifts for a documented pure $6s^2\rightarrow 6s6p$ transition~\cite{Zaal1}.  A King-Plot analysis~\cite{King} was performed using these two transitions to evaluate the specific mass shift of the 421-nm transition and the ratio of field-shift parameters. The electronic field-shift parameter $E_i$ is proportional to the change in electron probability density at the nucleus. If the excited-state configuration is the same for both transitions, the field-shift parameters should be approximately equal, and the slope must be close to unity.  From the slope of the King Plot line (Fig.~\ref{fig:KingPlot}), the ratio of electronic field-shift parameters $E_{421}/E_{457}=0.920(6)$ was obtained, indicating a close agreement with a configuration of $4f^{10}6s6p$ for the upper state. From the intercept and an assumed specific mass shift (SMS) of 7(8)~MHz for the reference transition~\cite{Zaal1}, the specific mass shift for the 421-nm transition was found to be $\delta \nu^{164-162}_{sms}=11(7)$ MHz.  This value is typical for transitions in dysprosium that do not involve changes in the f-electron shell~\cite{Zaal1}, and is consistent with a predicted value of $\delta \nu^{164-162}_{sms} = (0\pm0.5)\delta \nu^{164-162}_{nms}$ for an $s^2\rightarrow s\,p$ transition~\cite{sms}, where the normal mass shift ($\delta \nu^{164-162}_{nms}$) is $30$ MHz~\cite{Zaal1,Budker} for the 421-nm transition.  This is in contrast to transitions where the $f$-shell is rearranged. For example, a measurement of the specific mass shift for a radio-frequency $4f^{10}5d6s\rightarrow4f^95d^26s$ transition in dysprosium gives a value of $\delta \nu^{164-162}_{sms}=-516(50)$~MHz~\cite{BudkerThesis}.

\subsection{Hyperfine Coefficients}

The final values for the hyperfine-interaction coefficients are listed in Table~\ref{table:hyperfine}.  In certain cases of high-resolution spectroscopy, the magnetic-octupole term is required to achieve a good fit~\cite{BudkerThesis}, but the 421-nm spectrum was too densely packed to reliably determine this term and was consequently omitted in the final fitting procedure.  The ratios of $A^{163}/A^{161} = -1.3995(4)$ and $B^{163}/B^{161}=1.0558(4)$ are consistent with the ground state values~\cite{Childs} and other excited state values~\cite{Zaal2,BudkerThesis}, indicating no significant hyperfine anomaly within the experimental uncertainties.

\section{Conclusion}

We have discussed precision spectroscopic measurements of a potential cooling transition in atomic dysprosium at 421-nm.  Information on the hyperfine coefficients for both $I=5/2$ isotopes and isotope shifts for all isotopes have been measured.  King-Plot analysis of the isotope shifts against a pure $6s^2\rightarrow 6s6p$ transition was used to extract the specific mass shift and confirm the configuration of the upper state.  Work currently under way is dedicated towards identifying optical ``leaks" in this cycling transition and devising a hyperfine level repumping scheme for the cooling of the $I=5/2$ isotopes.

\begin{acknowledgements}
The authors are grateful to D. English for useful discussions, A.-T. Nguyen for extensive contributions to the design and construction of the atomic-beam apparatus, and J. Torgerson for help building the frequency doubler.  This work was supported by the Foundational Questions Institute (fqxi.org) and the UC Berkeley Committee on Research.
\end{acknowledgements}
\clearpage
\bibliography{421TransitionBib}

\end{document}